\documentstyle[12pt]{article}

\begin{document}

\title{A topological mechanism of discretization for the
electric charge}

\author{Antonio F. Ra\~nada\thanks{E-mail: fite113@sis.ucm.es}
and Jos\'e L. Trueba\thanks{E-mail: jltrueba@eucmos.sim.ucm.es}\\
Departamento
de F\'{\i}sica Te\'orica \\ Universidad Complutense \\ 28040 Madrid,
Spain }

\date{8 October 1997}

\maketitle

\begin{abstract}

We present a topological
mechanism of discretization, which gives for the fundamental electric
charge the value $q_0=\sqrt{\hbar c} =5.29\times 10^{-19}$ C, about 3.3
times the electron charge. Its basis is the following recently proved property of the
standard linear classical Maxwell equations: they can be obtained by
change of variables from an underlying topological theory, using two
complex scalar fields $\phi$ and $\theta$ the level curves of which
coincide with the magnetic and the electric lines, respectively.

\end{abstract} 

\newpage

The discretization of the electric charge is one of the most important
and intriguing laws of physics. However, the value of the fundamental
charge is obtained through experiments,
all the efforts to predict it --- or the fine structure constant $\alpha$
--- within a theoretical scheme having failed so far.

A model of electromagnetism proposed a few years ago \cite{Ran89,Ran92,Ran97} 
(to be called ``the topological model" from now on) uses the idea of force line as the basic dynamic element. It defines the magnetic and electric lines as the level curves of two complex scalar fields $\phi$ and $\theta$ so that each one is labelled by a complex number. The topology of
the force lines plays a very important role and has several curious consequences; this paper shows that
one of them is that the model contains
only electromagnetic fields which are coupled to electric charges equal
to $nq_0$, $n$ being an integer and $q_0=\sqrt{\hbar c} =5.29\times 10^{-19}$ C, about 3.3
times the electron charge, in such a way that, among the electric
lines converging
to or diverging from the charge, there are precisely $|n|$ whose label
is any
prescribed complex number.  The model is locally equivalent to Maxwell
standard theory, in the sense that any standard field is locally equal to one of the topological fields (so called electromagnetic knots), which shows that Maxwell equations are compatible with the existence of topological charges. However there is a difference of global character referring
to the way the fields surround the point at infinity (a natural way in the topological model, allowing only electromagnetic fields of finite energy in empty space, for instance.) All this shows the existence of a beautiful mathematical structure, unknown thus far and which is characteristic of Maxwell standard theory. In this paper, we consider this new idea.

This important law is usually stated by saying that
the electric charge of any particle is an integer multiple of a fundamental
value $e$, the electron charge, whose value in the International System of
Units is $e= 1.6 \times 10^{-19}$ C. The Gauss theorem allows a
different,
although fully equivalent, statement of this property: the electric flux
across any closed surface $\Sigma$ which does not intersect any charge
is always an integer multiple of $e$ (we will use here the
rationalized MKS system). This can be written as
 \begin{equation}
\int _{\Sigma} \omega = ne,
\label{1}
\end{equation}
where $\omega$  is the 2-form ${\bf E\cdot n}\, dS$, {\bf n} being a
unit vector orthogonal to the surface, {\bf E} the electric field
and $dS$ the surface element. We could as well write (\ref{1}) as
\begin{equation}
\int _{\Sigma}\  {*} {\cal F} =ne,
\label{2}
\end{equation}
${*}{\cal F}$ being the dual to the Faraday 2-form
${\cal F}=\frac{1}{2}F_{\mu\nu}dx^\mu\wedge dx^\nu$.
Stating in this way the discretization of the charge is interesting because
it shows a close similarity with the expression of the topological degree
of a map. Assume that we have a regular map $\theta$ of $\Sigma$ on a
2-sphere $S^2$ and let $\sigma$ be the normalized area 2-form in $S^2$.
It then happens that
\begin{equation}
\int _{\Sigma} \theta ^*\sigma =n,
\label{3}
\end{equation}
$\theta ^*\sigma$ being the pull-back of $\sigma$ and $n$ an integer
called the degree of the map, which gives the number of times that
$S^2$ is covered when one runs once through $\Sigma$ (equal to the
number of points in $\Sigma$ in which $\theta$ takes any prescribed
value). Note that $\theta
^*$ in (\ref{3}) indicates pull-back by the map $\theta$ and must not be
mistaken for the complex conjugate of $\theta$, which will be written
$\bar \theta$.

The comparison of (\ref{2}) and (\ref{3}) shows that there is a close
formal similarity between the dual to the Faraday 2-form and the pull-back
of the area 2-form of a sphere $S^2$. It can be expressed in this way.
Let an electromagnetic field be given, such that its form ${*}{\cal F}$
is regular except at the positions of some point charges. Let a map
$\theta :R^3\mapsto S^2$ be also given, which is regular except at some
point singularities where its level curves converge or diverge. It happens
then that equations (\ref{2}) and (\ref{3}) are simultaneously satisfied
for all the closed surfaces $\Sigma$ which do not intersect any charge
or singularity.

This means that the electric charge will be automatically and
topologically
discretized in a model in which these two forms --- ${*}{\cal F}$ and
$\theta ^*\sigma$ --- are proportional, the fundamental charge being
equal to the proportionality coefficient and the number of fundamental
charges in a volume having then the meaning of a topological index.

A way to achieve such a model is the following. Let a complex scalar
field
$\theta ({\bf r},t)$ be given. Via stereographic projection, the complete
complex plane can be identified with the sphere $S^2$, so that $\theta$
can be interpreted as giving a map $\theta:R^3\mapsto S^2$ at any
time.
Suppose now that there is an electromagnetic field such that the
dual to
its Faraday form ${*}{\cal F}$ and the pull-back $\theta^{*} \sigma$ are
proportional, ${*}{\cal F} = \lambda \theta^{*} \sigma$. As the
dimensions
of $\lambda$ are square root of action times velocity, this can be
written as
 \begin{equation}
{*} {\cal F} =\sqrt{a} \, \theta^*\sigma ,
\label{5}
\end{equation}
where $a$ is a normalizing constant with dimensions of action times velocity.
It turns out then that
\begin{equation}
{*}\!{\cal F}= \sqrt{a}\,\theta ^*\sigma
=\frac{\sqrt{a}}{2\pi i}\,
\frac{d\theta \wedge d{\bar \theta}}{(1+{\bar \theta} \theta )^2},
\label{4}
\end{equation}
the dual to the Faraday tensor being then
\begin{equation}
{*} F_{\mu\nu}= \frac{\sqrt{a}}{2\pi i}\,
\frac{\partial _\mu \theta \partial _\nu {\bar \theta}
- \partial _\nu \theta \partial _\mu {\bar \theta}}{(1+{\bar \theta}
\theta )^2}.
\label{6}
\end{equation}
Note that  the electric field is
${\bf E}=\sqrt{a}\,(2\pi i)^{-1}(1+\bar{\theta}\theta )^{-2}
\nabla \bar{\theta} \times \nabla \theta$ (because of (\ref{6})), so
that the electric lines are the level curves of $\theta$.
The degree of the map $\Sigma\mapsto S^2$ induced by $\theta$ is given
by (\ref{3}); it turns out therefore that
\begin{equation}
\int _{\Sigma} \ {*}{\cal F} = n\sqrt{a}.
\label{7}
\end{equation}
As this is equal to the charge $Q$ inside $\Sigma$, it does happen that
$Q=n\sqrt{a}$, what implies that there is then a fundamental charge
$q_0=\sqrt{a}$,
the degree $n$ being the number of fundamental charges inside $\Sigma$.
This gives a topological interpretation of $n$.

It is easy to understand that $n=0$ if $\theta$ is regular in
the interior of $\Sigma$. This is because each level curve of $\theta$
({\em i. e.} each electric line) is labelled by its value along it --- a
complex number --- and, in the
regular case, any one of these lines enters into this interior as many
times as it goes out of it. But assume that $ \theta$ has
a singularity at point $P$, from which the electric lines diverge or to
which
they converge. If $\Sigma$ is a sphere around $P$, we can identify $R^3$
except $P$ with $\Sigma \times R$, so that the induced map
$\theta :\Sigma \mapsto S^2$ is regular. In this case, $n$ need not
vanish
and is equal to the number of times that $\theta$ takes any prescribed
complex value in $\Sigma$, with due account to the orientation. Otherwise
stated, among the electric lines diverging from or converging to $P$,
there are $|n|$ whose label is equal to any prescribed complex number.

This shows a recipe for constructing a model with topological
discretization of the charge: just guarantee that (\ref{5}) (or
equivalently (\ref{6})) will be satisfied. Opportunely enough, it turns
out that the topological model above mentioned is of this kind (see \cite{Ran89,Ran92,Ran97,Ran91} and
references therein; the same ideas have inspired 
a model of ball lightning \cite{Ran96}), which is
based on the dynamics of the force lines and makes use of two scalar
fields $\phi$
and $\theta$, interpreted as maps $S^3\mapsto S^2$ to define the
electromagnetic
radiation fields through (\ref{5})-(\ref{6}) and the analogous relations
\begin{equation}
{\cal F} =-\sqrt{a} \, \phi ^*\sigma ,
\label{9}
\end{equation}
\begin{equation}
F_{\mu\nu}= \frac{\sqrt{a}}{2\pi i}\,
\frac{\partial _\mu {\bar \phi} \partial _\nu \phi
- \partial _\nu {\bar \phi} \partial _\mu \phi}{(1+{\bar \phi} \phi
)^2},
\label{10}
\end{equation}
As the magnetic field is
${\bf B}=-\sqrt{a}\, (2\pi i)^{-1}(1+\bar{\phi}\phi )^{-2}
\nabla \bar{\phi} \times \nabla \phi $,
the magnetic and electric lines are the level curves of
$\phi$ and $\theta$, respectively.

To understand better this mechanism of discretization, let us take the
case of a Coulomb potential \cite{Ran91,Ran97}, ${\bf E} = Q {\bf r}/(4\pi
r^3)$, ${\bf B} =0$.  The corresponding scalar is then
\begin{equation}
\theta = \tan{\left(\frac{\vartheta}{2}\right)} \: {\mbox {exp}} \:
\left( i \frac{Q}{\sqrt{a}} \varphi \right) ,
\label{sca}
\end{equation}
where $\varphi$ and $\vartheta$ are the azimuth and the polar angle.
The scalar (\ref{sca}) is well defined only if $Q = n \sqrt{a}$,
$n$ being an integer.  The lines diverging from the charge are labelled
by the corresponding value of $\theta$, so that there are $|n|$ lines
going in or out of the singularity and having any prescribed complex
number
as their label. If $n=1$, it turns out that $\theta = (x+iy)/(z+r)$.

This mechanism has a very curious aspect: it does not apply to the
source but to the electromagnetic field itself.  This is surprising; one
would expect that the topology should operate restricting the fields of
the charged particles.  However, in this model, it is the field who
mediates the force the one which is submitted to a topological condition.

These two scalars are assumed to
represent maps $S^3\mapsto S^2$, which are regular except for
singularities at the position of point charges. Consider first the case
of empty space, without charges. Four properties of the topological
model must be emphasized.

(i) The maps $\phi$ and $\theta$ are dual the
one to the other in the sense that (\ref{5})-(\ref{9}) are verified
simultaneously (in a more formal notation, this is written $*(\phi
^*\sigma)=-\theta
^*\sigma$, $*$ being  the Hodge or duality operator). Surprisingly,
their
mere existence implies that the electromagnetic tensor and its dual
obey necessarily the equations
(with $F_{\mu\nu}$ and $*F_{\mu\nu}$ given by (\ref{6}) and (\ref{10}))
\begin{equation}
\partial ^\mu F_{\mu\nu}(\bar{\phi},\phi )=0,\;\;\;\partial ^\mu {*}\!
F_{\mu\nu}(\bar{\theta},\theta )=0,
\label{a1}
\end{equation}
which can be obtained also by an action principle with
the usual Lagrangian density $-F_{\mu\nu}F^{\mu\nu}/4$, but taking the
two scalars as fundamental
fields \cite{Ran92,Ran97}. This can be stated also as follows. Let us define the product map $\chi =\phi \times \theta :S^3\mapsto S^2\times S^2$, so that ${\cal V}=\chi ^*(\sigma \wedge \sigma )=\phi ^*\sigma \wedge \theta ^*\sigma $ is the pull-back to $S^3$ of the volume form in $S^2\times S^2$. It happens then that $a\,{\cal V}/2=-{\cal F}(\phi)\wedge
{*}{\cal F}(\theta )/2=-F_{\mu\nu}F^{\mu\nu}d^4x/4$, if the scalars satisfy the above duality condition. Surprisingly enough, it turns out that the action
\begin{equation}
{\cal S}=\frac{a}{2c}\int {\cal V},
\label{a2}
\end{equation}
takes a stationary value for any pair of maps $\phi ,\theta$.

Note that (\ref{a1}) are highly nonlinear in
the scalars but become exactly the linear Maxwell
equations in the fields $F_{\mu\nu}$ and ${*}\!F_{\mu\nu}$. In this
sense, the Maxwell equations are the exact linearization (by change of
variables, not by truncation!) of a nonlinear
theory with topological properties, in which the force lines coincide
with the level curves of two scalar fields. The model gives thus a
line-dynamics.

(ii) To require that the scalars give regular maps $S^3\mapsto
S^2$ is equivalent to impose on them a condition of compactification:
that they have only one value at infinity. This is what allows
compactifying the space $R^3$ to $S^3$ and
has an important consequence: the corresponding
Hopf indices are two topological constants of the motion,
equal to the linking numbers of any pair of electric or magnetic lines.
It
happens that these two numbers are equal  (say to the integer $m$)
and equal
also, up to the factor $a$, to the common value of the magnetic
and electric helicities
 \begin{equation}
h_{m}= \int {\bf A\cdot B}\,d^3r=m a,\;\;\;h_{e}= \int {\bf
C\cdot E}\,d^3r=m a,
\label{11}
\end{equation}
where {\bf A} and {\bf C} are vector potentials for {\bf B} and {\bf E}
and the integrals extend to all $R^3$ (see
\cite{Ran92,Mof69,Mof92,Mar96} for discussions
on the idea of helicity).  The corresponding waves have been called {\em
electromagnetic knots}  because any pair of magnetic lines and
any pair of electric lines is a link with the same linking number $n$.
In references \cite{Ran90,Ran95,Ran97} the explicit expressions for
families of such knots are given.

(iii) The electromagnetic knots have
the following nice property: any standard radiation electromagnetic
field defined in
a bound domain of spacetime is locally equal to an electromagnetic knot,
except on a zero measure set (for the proof, see section 4 of
\cite{Ran92} and section 2 of \cite{Ran97}). In this precise sense,
it can be said that the sets
of the electromagnetic knots and of the standard radiation fields
coincide.
 As any standard field is locally equal to the sum
of two radiation
fields, the topological model is locally equivalent to Maxwell standard
theory, although they are different globally.  This means that one can
not distinguish the one from the other by looking only locally as it is
done in most experiments.  The only difference is of global character
and refers to the way in which the fields surround the infinity. (Note
incidentally that the Bohm-Aharonov effect which has nonlocal character
requires topological considerations.)

(iv) It turns out that the electromagnetic knots verify
\begin{equation}
m a  = \hbar c(N_R-N_L) ,
\label{12}
\end{equation}
where $m$ is the linking number of any pair of magnetic or electric
lines, and
$N_R$, $N_L$ are the classical expressions for the numbers of right
and left-handed photons ({\em i. e.} $N_R=\int \bar{a}_Ra_R d^3k$,
$N_L=\int \bar{a}_La_L d^3k$, the functions $\bar{a}_{R,L}, a_{R,L}$
being here Fourier
transforms of the classical vector potential {\bf A}, the {\em c-}number
fields which are interpreted in QED as creation and annihilation operators
for right and left polarization photons. See \cite{Ran97} and references
therein for the
proof). It must be stressed that the right hand side of  (\ref{12}) is fully
meaningful and is well defined as a classical quantity.

Equation (\ref{12}) relates in a simple an elegant way the two meanings
of the word
helicity, referring to the wave and particle aspects of the field. At left,
the wave helicity $m a$ characterizes the way in
which the force lines curl
themselves the ones around the others; at right, $\hbar c$ times the particle helicity
$N_R-N_L$, to which right- and left-handed photons contribute with $+1$
and $-1$, respectively. They are equal. Suprisingly this relation
arises in a classical context
because of topological reasons. The consequence is that $N_R-N_L$ is
topologically discretized in the topological model, even in the fields
are
{\em c}-numbers, suggesting that it could give a classical limit with the
right normalization. This suggests taking $a=\hbar c$, so that $N_{R} - N_{L}
=
m$. This is surprising: $N_{R} - N_{L}$ is then equal to the linking number
of
the force lines! (and as such it is always an integer number and can
take any integer value).

These properties may seem strange in a theory which uses the linear Maxwell equations. However, there is no contradiction, since the electromagnetic fields verifying (\ref{6}) and (\ref{10}) form a nonlinear subset of the vector space of the solutions of Maxwell equations, in spite of its set being locally equivalent to the set of the standard fields. This property has been called hidden nonlinearity \cite{Ran92,Ran95}.

All the necessary details about this topological model  are explained in
the above quoted references \cite{Ran89,Ran90,Ran91,Ran92,Ran95,Ran97}
(as a sign of the current interest in linked and knotted configurations
of classical fields, see reference \cite{Fad97}).

The previous arguments, four properties, and equations (\ref{a1}),
(\ref{11}) and (\ref{12}) refer
to the case of empty space, where the maps $\phi$, $\theta$ are
everywhere regular.
To include point charges, the scalar $\theta$ must have singularities
where its level curves ({\em i. e.} the electric lines) converge or
diverge. As explained above the value of the charge must be then
topologically discretized. With $a=\hbar c$, the fundamental charge is
\begin{equation}
q_0=\sqrt{\hbar c},
\label{12b}
\end{equation}
(in the MKS system) which is about $3.3$ times the electron charge. In
the ISU, this is $q_0=\sqrt{\hbar c\epsilon _0}=5.29\times 10^{-19}$ C,
and in natural units $q_0=1$.

To summarize, in the
topological model explained in \cite{Ran89,Ran90,Ran91,Ran92,Ran95,Ran97},
the magnetic and electric lines are described as the level curves of
two complex scalar fields $\phi$ and $\theta$. As a consequence of the
topology of these
lines, some integer numbers characterize the electromagnetic fields.
In empty space, the fields are called electromagnetic knots, since any
pair
of magnetic or electric lines are linked with the same linking number
$m$,
the magnetic and electric helicities being both $m\hbar c$. It turns out
that $m=N_R-N_L$, $N_R, N_L$ being the classical expressions which are
interpreted in QED as numbers of right- and left-handed photons.
Suprisingly for a topological model, it is locally equivalent to Maxwell
theory (because any standard field is locally equal to a field of the
model, generated by two scalars), the difference being of global
character and referring to the way the fields
behave around the point at infinity. Furthermore, the electromagnetic
fields
can only be coupled to point charges which are integer multiple of
the fundamental charge $q_0=\sqrt{\hbar c}$. Note that the same
discretization mechanism would apply to the hypothetical magnetic
charges (located at singularities of $\phi$), their fundamental value
being also $q_0=\sqrt{\hbar c}$.

Monopoles can be included, therefore, in this model in a natural and
simple way, but with a magnetic charge different from the well known
Dirac value $g=2\pi /e$.  To be specific, one has in natural units
$e=0.3028$ and $g=20.75$, so that $e<q_0<g$.  In other words,
the fundamental charge $q_0$ of this model is bigger than the electron
charge $e$, but smaller than the Dirac monopole $g$.  Maybe it should be
mentioned that, as the vacuum is dielectric and paramagnetic,
the quantum corrections due to the sea of virtual pairs should decrease
the electric charge but increase the magnetic one (in other words, the
observed electric charge must be smaller than the bare one, the opposite
being true for the magnetic charge)

This topological model shows thus a striking electromagnetic duality, 
since the  electric and the magnetic charge have the same value 
$\sqrt{\hbar c}$ at the classical level (at which this model applies 
in its present form and where there are no vacuum polarization effects) 
and, moreover, they are inverse to one another, in natural units. 
Consequently, if the sea of virtual pairs is included, the observed 
electric and magnetic charges must be smaller and bigger than 
$\sqrt{\hbar c}$, respectively, in agreement with current knowledge, 
since the Dirac monopole must be understood as a dressed magnetic charge.

\bigskip

\bigskip

{\bf Acknowledgements.}
We are grateful to Profs. Michael V. Berry, Alberto Ibort, Jos\'e M.
Montesinos and Mario Soler for discussions end encouragement, and
to Prof. Alfredo Tiemblo for discussions and hospitality to A. F. R. at
IMAFF, CSIC, Madrid. This paper has been partially supported by DGICYT 
under grant PB95-0383.


\bigskip
\bigskip

\begin{thebibliography}{123456}
\bibitem{Ran89} A. F. Ra\~nada, A topological theory of the
electromagnetic field, Lett. Math. Phys. 18 (1989) 97-106.
\bibitem{Ran92} A. F. Ra\~nada, Topological electromagnetism, 
J. Phys. A: Math. Gen. 25 (1992) 1621-41.
\bibitem{Ran97} A. F. Ra\~nada and J. L. Trueba, Two properties of
electromagnetic knots, Phys. Lett.  A 235 (1997) 25-33.
\bibitem{Ran91} A. F. Ra\~nada, A topological model of electromagnetism: 
Quantization of the electric charge, An. Fis. (Madrid) A 87 (1991) 55-59.
\bibitem{Ran96} A. F. Ra\~nada and J. L. Trueba, Ball lightning an
electromagnetic knot?, Nature 383 (1996) 32.
\bibitem{Mof69} H. K. Moffatt, The degree of knottedness of tangled
vortex lines, J. Fluid. Mech. 35 (1969) 117-129.
\bibitem{Mof92} H. K. Moffatt and R. L. Ricca, Helicity and the
C\u{a}lug\u{a}reanu invariant, Proc. R. Soc. Lond. A 439 (1992) 411-429.
\bibitem{Mar96} G. E. Marsh, Force-free magnetic fields
(World Scientific, Singapore, 1996).
\bibitem{Ran90} A. F. Ra\~nada, Knotted solutions of the Maxwell
equations in vacuum, J. Phys. A:Math. Gen 23 (1990) L815-820.
\bibitem{Ran95} A. F. Ra\~nada and J. L. Trueba, Electromagnetic
knots, Phys. Lett. A 202 (1995) 337-342.
\bibitem{Fad97} L. Faddeev and A. J. Niemi, Stable knot-like
structures in classical field theory, Nature 387 (1997) 58.

\end{thebibliography}
\end{document}